\documentclass[twocolumn,aps,pra,superscriptaddress,amsmath,amssym,floatfix,longbibliography]{revtex4-1}
\usepackage[T1]{fontenc}
\usepackage[utf8]{inputenc}
\usepackage{graphicx}% Include figure files
\usepackage{dcolumn}% Align table columns on decimal point
\usepackage{bm}% bold math
\usepackage{mathrsfs}
\usepackage{amsmath}
\usepackage{float}
\usepackage{soul}
\usepackage{lipsum}
\usepackage{physics}
\usepackage{dsfont}
\usepackage{color}
\usepackage{tikz}
\usepackage{enumitem}
\usepackage{comment}

\usepackage[colorlinks=true,linkcolor=blue,citecolor=blue,filecolor=green,urlcolor=blue]{hyperref}

\def\la{\langle}
\def\ra{\rangle}

\def\be{\begin{equation}}
\def\ee{\end{equation}}

\newcommand\e{\mathrm e}
\usepackage[squaren,cdot]{SIunits}

\begin{document}

\newcommand{\bigjprob}{{\mathcal{P}}}
\newcommand{\bigprob}{_{\bm{q}_F}{\mathcal{P}}_{\bm{q}_I}}
\newcommand{\cum}[1]{\llangle #1 \rrangle}       					% Cumulant
\newcommand{\vop}[1]{\vec{\bm #1}}
\newcommand{\opt}[1]{\hat{\tilde{\bm #1}}}
\newcommand{\vopt}[1]{\vec{\tilde{\bm #1}}}
\newcommand{\td}[1]{\tilde{ #1}}
\newcommand{\mean}[1]{\la#1\ra}                  					% Mean
\newcommand{\cmean}[2]{ { }_{#1}\mean{#2}}       				% Conditioned Mean
\newcommand{\pssmean}[1]{ { }_{\bm{q}_F}\mean{#1}_{\bm{q}_I}}
\newcommand{\ipr}[2]{\left\la#1\mid#2\right\ra}            				% Inner Product
\newcommand{\opr}[2]{\ket{#1}\bra{#2}}           					% Outer Product
\newcommand{\pr}[1]{\opr{#1}{#1}}                					% Projector 
\newcommand{\Trd}[1]{\text{Tr}_d(#1)}            					% Partial trace over detector
\newcommand{\Trs}[1]{\text{Tr}_s(#1)}            					% Partial trace over system
\newcommand{\intd}[1]{\int \! \mathrm{d}#1 \,}
\newcommand{\fullint}{\iint \! \mathcal{D}\mathcal{D} \,}
\newcommand{\drv}[1]{\frac{\delta}{\delta #1}}
\newcommand{\partl}[3]{ \frac{\partial^{#3}#1}{ \partial #2^{#3}} }		% Partial Derivative
\newcommand{\smpartl}[4]{ \left( \frac{\partial^{#3} #1}{ \partial #2^{#3}} \right)_{#4}}
\newcommand{\smpartlmix}[4]{\left( \frac{\partial^2 #1}{\partial #2 \partial #3 } \right)_{#4}}
\newcommand{\limit}[2]{\underset{#1 \rightarrow #2}{\text{lim}} \;}
\newcommand{\funcd}[2]{\frac{\delta #1}{\delta #2}}
\newcommand{\funcdiva}[3]{\frac{\delta #1[#2]}{\delta #2 (#3)}}
\newcommand{\funcdivb}[4]{\frac{\delta #1 (#2(#3))}{\delta #2 (#4)}}
\newcommand{\funcdivc}[3]{\frac{\delta #1}{\delta #2(#3)}}
\newcommand{\tkw}[1]{\textcolor{red}{#1}}
\definecolor{dgreen}{RGB}{30,130,30}
% Use the \preprint command to place your local institutional report
% number in the upper righthand corner of the title page in preprint mode.
% Multiple \preprint commands are allowed.
% Use the 'preprintnumbers' class option to override journal defaults
% to display numbers if necessary
%\preprint{}

%Title of paper
\title{%Note about a simplified model of the fluxon engine
%
%Cyclic fluxon quantum refrigerators
%A cyclic fluxon quantum fridge
Cyclic superconducting quantum refrigerators using guided fluxon propagation
}
	\author{Tathagata Karmakar}
	\email{tkarmaka@ur.rochester.edu}
	\affiliation{Department of Physics and Astronomy, University of Rochester, Rochester, NY 14627, USA}
	\affiliation{Center for Coherence and Quantum Optics, University of Rochester, Rochester, NY 14627, USA}
	\affiliation{Institute for Quantum Studies, Chapman University, Orange, CA 92866, USA}
\author{Étienne Jussiau}
\email{ejussiau@ur.rochester.edu}
\affiliation{Department of Physics and Astronomy, University of Rochester, Rochester, NY 14627, USA}
\affiliation{Center for Coherence and Quantum Optics, University of Rochester, Rochester, NY 14627, USA}
\affiliation{Institute for Quantum Studies, Chapman University, Orange, CA 92866, USA}
\author{Sreenath K.~Manikandan}
	\email{sreenath.k.manikandan@su.se}
	\affiliation{Nordita,
KTH Royal Institute of Technology and Stockholm University,
Hannes Alfv\'{e}ns v\"{a}g 12, SE-106 91 Stockholm, Sweden}
\author{Andrew N.~Jordan}
\email{jordan@chapman.edu}
\affiliation{Institute for Quantum Studies, Chapman University, Orange, CA 92866, USA}
\affiliation{Department of Physics and Astronomy, University of Rochester, Rochester, NY 14627, USA}
\affiliation{Center for Coherence and Quantum Optics, University of Rochester, Rochester, NY 14627, USA}

\date{\today}

\begin{abstract}

We propose cyclic quantum refrigeration in solid-state, employing a gas of magnetic field vortices in a type-II superconductor---also known as fluxons---as the cooling agent. Refrigeration cycles are realized by envisioning a racetrack geometry consisting of both adiabatic and isothermal arms, etched into a type-II superconductor.  The guided propagation of fluxons in the racetrack is achieved by applying an external electrical current, in a Corbino geometry,
through the sample. A gradient of magnetic field is set across the racetrack allowing one to adiabatically cool down and heat up the fluxons, which subsequently exchange heat with the cold, and hot reservoirs, respectively. We characterize the steady state of refrigeration cycles thermodynamically for both $s-$wave and $d-$wave pairing symmetries, and present their figures of merit such as the cooling power delivered, and the coefficient of performance. Our cooling principle can offer significant cooling for on-chip micro-refrigeration purposes, by locally cooling below the base temperatures achievable in a conventional dilution refrigerator. We estimate $\unit{10}{\nano\watt\per\milli \meter\squared}$ of cooling power per unit area under typical operating conditions. 
Integrating the fluxon fridge to quantum circuits can enhance their coherence time by locally suppressing thermal fluctuations, and improve the efficiency of single photon detectors and charge sensors.%\tkw{**maybe one last sentence about where it might be useful?**}
%This note gives my calculations about a simplified model of the fluxon fridge
\end{abstract}

\maketitle
\section{Introduction} \label{sec:intro}

 %In an external magnetic field gradient 
Quantum thermal machines have attracted increasing amounts of attention not only because of their fundamental importance in developing the field of quantum thermodynamics \cite{benenti2017fundamental}, but also because of the practical importance of controlling thermal transport in cryogenic environments \cite{sothmann2014thermoelectric}. Long-standing material science problems of low thermoelectric efficiencies can be overcome via energy-structured transport properties between coupled conductors \cite{mahan1996best}. In the field of mesoscopic physics, extensive research has been carried out to investigate thermoelectric devices \cite{van1992thermo}, heat engines \cite{jordan2013powerful}, thermometers \cite{giazotto2006opportunities}, heat diodes \cite{sanchez2015heat}, heat transistors \cite{yang2019thermal}, and refrigerators \cite{manikandan2020autonomous}. These devices are primarily focused on electrons as the charge and heat carriers based on quantum dots \cite{sanchez2011optimal}, quantum wells \cite{sothmann2013powerful}, quantum point contacts \cite{sothmann2012rectification} and superlattices \cite{choi2015three}. However, research focusing on other heat carriers including magnons \cite{sothmann2012magnon}, phonons \cite{saira2007heat}, photons \cite{henriet2015electrical}, and superconducting degrees of freedom \cite{hofer2016quantum} has also advanced quickly. Not only a theoretical discipline but ground-breaking experiments have also been performed, realizing refrigerators \cite{prance2009electronic} and heat engines leveraging the sharp energy transmission features of resonant tunneling quantum dots \cite{jaliel2019experimental}, current rectification properties of electron cavities coupled with quantum point contacts to Ohmic contacts \cite{roche2015harvesting,hartmann2015voltage}, as well as driven superconductors to control heat currents \cite{senior2020heat,fornieri2017towards}. Together, these experiments have demonstrated high efficiency thermal machines operating on quantum principles.
 
 \begin{figure}[tbh]
    \centering
% \begin{tikzpicture}
%\node[] at (0,0) {\includegraphics[width=\linewidth]{Racetrack1.pdf} };
%\node[] at (1.7,2.1)
%{\color{black} $\odot$};
%\node[] at (2,2.1)
%{\color{black} $\odot$};
%\node[] at (2,2.4)
%{\color{black} $\odot$};
%\node[] at (1.7,2.4)
%{\color{black} $\odot$};
%\node[] at (-2.3,2.1)
%{\color{black} $\odot$};
%\node[] at (-2,2.1)
%{\color{black} $\odot$};
%\node[] at (-2.2,2.5)
%{\color{black} $H_\mathrm L$};
%\node[] at (1.3,2.5)
%{\color{black} $H_\mathrm R$};
%\node[] at (1.3,1.8)
%{\color{black} $T_\mathrm R (\phi_\mathrm R)$};
%\node[] at (1.4,-2)
%{\color{black} $\tilde{T}_\mathrm C (\tilde{\phi}_\mathrm C)$};
%\node[] at (-2,-2)
%{\color{black} $T_\mathrm L (\phi_\mathrm L)$};
%\node[] at (-2,1.8)
%{\color{black} $\tilde{T}_\mathrm H (\tilde{\phi}_\mathrm H)$};
% \end{tikzpicture}   
\includegraphics[width=\linewidth,trim={10 40 10 40},clip]{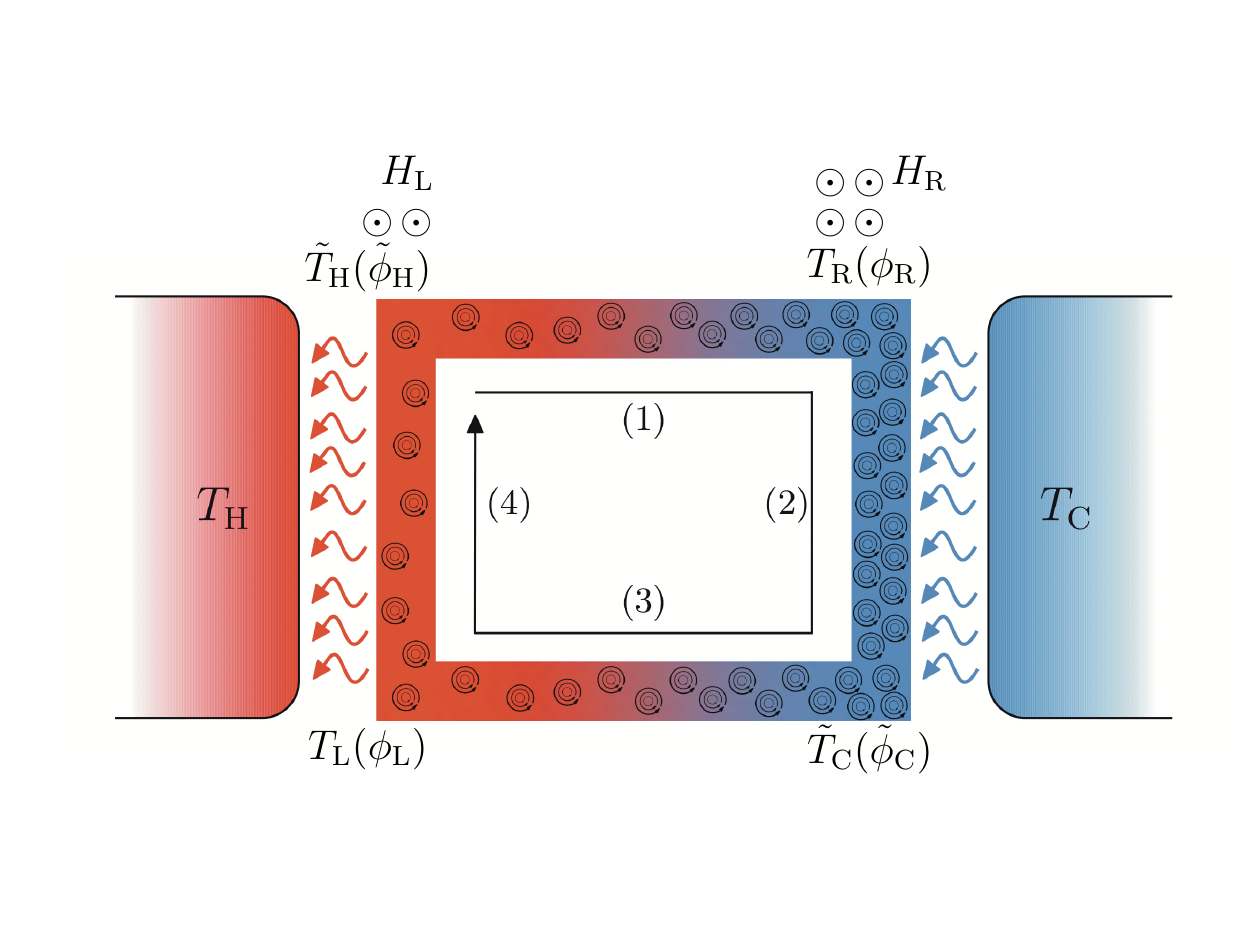} 
    \caption{Schematic of the type-II superconducting refrigerator geometry with two arms in contact with hot ($T_\mathrm H$) and cold ($T_\mathrm C$) reservoirs. A magnetic field (out of the page) gradient from arm (4) to arm (2) causes a gradient in the vortex density. An external current (not shown) flows outward in the Corbino geometry, driving the vortices along the arrow shown. The temperatures at the four corners represent the temperatures of the fluxons at the end of each stroke (see sections \ref{model}, \ref{dynamics}). The values in the bracket represent the corresponding temperatures scaled with respect to that of the hot reservoir (see section \ref{dimensionless}). }
    \label{geometry}
\end{figure}

Here, we consider another type of heat carrier - the fluxon. In a type-II superconductor, magnetic flux quanta can pierce the entropy-free Cooper-paired superconducting state, to produce an island of normal electrons in the core of each fluxon~\cite{abrikosov1957magnetic,abrikosov2004nobel}. We utilize this fluxon as a bucket of entropy, shuttling it back and forth between hot and cold reservoirs. To make a thermodynamic %engine 
cycle, our other control parameter is the local magnetic field. By magnetizing or demagnetizing a fluxon, the small electron gas inside the fluxon is cooled or heated. When confined to a racetrack geometry, Fig.~\ref{geometry}, a fluxon undergoes successive heating or cooling via the magnetic (de)magnetization from a gradient (out of plane) magnetic field  \cite{pioro2008electrically,wu2014two,kim2020long,banerjee2014reversible}, together with exposure to a hot or cold thermal reservoir. The remaining piece of the thermodynamic cycle %engine 
is the motive force needed to drive the fluxons around the racetrack. This is provided by the Lorentz force~\cite{tinkham2004introduction,embon2017imaging}, applied via a current bias applied in a Corbino geometry.

 % Mesoscopic samples of superconductors  give rise to fascinating emergent phenomena involving magnetic field-lines that penetrate the sample. This regime of physics has also been made accessible to experimentalists, thanks to the valiant efforts and recent progresses made in low-temperature physics, where controlled manipulation of local magnetic field gradients has been achieved~\cite{pioro2008electrically,wu2014two,kim2020long,banerjee2014reversible}. If the applied magnetic field is above the critical field, most of the elemental superconductors\footnote{with the exception of Niobium, Technetium and Vanadium} undergo a type-II transition through an intermediate state comprising of alternating regions of normal and superconducting phases. In contrast, it is now well-known that a 

 %Given these insights on the dynamics of fluxons, one particularly fascinating, and yet challenging application that we intend to explore in the present manuscript pertains to achieving quantum-enhanced cooling in mesoscopic scales, employing fluxons. Opportunities for cooling in mesoscopic scales are vast, given the  increasing number of candidate physical systems with potential applications ranging from quantum information processing~\cite{saffman2010quantum,blais2021circuit,burkard2021semiconductor}, quantum enhanced sensing~\cite{degen2017quantum,pirandola2018advances}, to exploring new avenues of fundamental physics in two-dimensions~\cite{sarma2011electronic,gu2018colloquium,von1986quantized}, and beyond. 
 The thermal machine briefly described above may be used as a refrigerator to cool the cold reservoir. We focus on the physics of refrigeration using this concept of a cyclic superconducting refrigerator.  Finding new cooling mechanisms at low temperature that do not rely on liquid Helium-3, a precious resource, is an outstanding challenge to the low temperature physics community. 
 %Broadly, the limitations to cooling are imposed by achievable ambient temperatures in dilution refrigerators, which has recently reached the single-digit, milli-Kelvin figures~\cite{giazotto2006opportunities}. Most dilution refrigerators use isotopes of helium for cooling, while some use magnetic calorimetry as their working principle, where magnetic samples, for example, paramagnetic salts are cooled down when they are demagnetized adiabatically~\cite{zu2022development,lounasmaa1979dilution}. 
 Principles of superconductivity offer another cooling paradigm, where the adiabatic magnetization of superconductors has been predicted to produce a cooling effect~\cite{keesom1934further,mendelssohn1934magneto,yaqub1960cooling}. Efforts have been made to find realistic applications for this approach since the early days of superconductivity, especially using conventional (type-I) superconductors ~\cite{dolcini2009adiabatic}. %whereas the initial temperature $T_{i}\rightarrow 0$, the achievable final temperatures scales cubically upon magnetization, i.e., $T_{f}\propto T_{i}^{3}$~\cite{dolcini2009adiabatic}.  
 Recently, it has also been proposed that cyclic quantum refrigerators can be conceptualized with type-I superconductors as the working substance, which can lead to practical quantum device implementations in solid-state~\cite{manikandan2019superconducting}.     
 % In the present article, we focus on a different possibility of cooling with adiabatic magnetization---demagnetization cycles,  using type-II superconductors as the working substance. More precisely, here we are interested in the physics of a fluxon refrigerator.  The geometry is a Corbino one, with voltage applied between outer and inner contacts, resulting in an outward electrical current $J$. 
%Similar setups have been used to generate and control thermal bias in the quantum Hall regime~\cite{real2022controlled}.  %A collection of fluxons is then put into a circular motion around the geometry, forming a `racetrack'.  Hot and cold thermal reservoirs are in contact with two edges of the sample, and a magnetic field gradient is applied to increase from the hot to the cold side of the sample. See Fig.~\ref{geometry}.

In contrast to type-I superconductors, the majority of non-elemental superconductors undergo a type-II phase transition into a mixed state with the magnetic field-lines forming flux vortices
~\cite{abrikosov1957magnetic,abrikosov2004nobel}, with each vortex, or fluxon,  carrying a quantum ($\hbar/2e$)  of magnetic flux. These flux vortices are known to organize into characteristic lattice structures, known as Abrikosov lattices~\cite{abrikosov1957magnetic,rosenstein2010ginzburg,maniv2001vortex,sonier2000musr}.   
 %The collective behavior of fluxons are extremely fascinating, given that each fluxon carries a flux quantum ($\hbar/2e$) unit of magnetic flux, and they more-or-less behave as emergent particles.  
 Individual fluxons have recently been proposed as information bits for efficient random-access memory devices~\cite{golod2015single}.  Fluxons also interact with spin-waves in superconductor/ferromagnetic hetero-junctions, suggesting opportunities for new avenues of hybrid electronic devices in the nano-scale~\cite{dobrovolskiy2019magnon}.  %Furthermore, guided propagation of fluxons is also possible as fluxons experience a Lorentz force in an external applied current density~\cite{tinkham2004introduction,embon2017imaging}. 
 The refrigerator proposed here further pushes the frontiers of fluxon based quantum technologies to solid-state integrable quantum devices which can offer substantial cooling below ambient base temperatures in cryogenic environments.  %Such progresses made in recent years in in low-temperature physics employing fluxons. %\tkw{**some concrete examples**}   

This article is organized as follows. Section \ref{model} describes the refrigerator and provides an account of the relevant cycles. We also calculate the heat exchanged, the cooling power delivered, and the coefficient of performance for $s$-wave and $d$-wave superconductor based refrigerators. Section \ref{dynamics} provides a phenomenological description of fluxon propagation and heat exchange with the reservoirs for $s$-wave superconductors. In light of this description, we characterize the performance of the refrigerator in section \ref{quanti}. Lastly, 
 we discuss our findings in section \ref{discussions} and conclude in section \ref{conclusions}. 

\section{\label{model}The model}

We now expand on the description of our fluxon heat engine, described in the introduction.  An insightful analogy of this heat engine can be made to domestic refrigerators operating based on the principles of free expansion and compression of non-ideal gases, cyclically moving through a cooling system. The fluxons moving through the Corbino racetrack geometry behave similarly, where the density of the ``fluxon gas" in different regions is controlled by the local magnetic induction. Fluxons are colder in regions of high magnetic field as opposed to regions of lower magnetic field, owing to adiabatic conditions assumed. The effective description for a cooling cycle with adiabatic and isothermal arms is presented in greater detail in the subsequent sections, and is comparable to the well-known Otto cycles~\cite{mozurkewich1982optimal,kosloff2017quantum,lior1988second}.  We consider fluxons in type-II superconductors of both $s-$wave and $d-$wave pairing symmetries, and provide a complete thermodynamic characterization, as well as discuss the laws of thermodynamics for the cyclic quantum refrigerator. 

\subsection{Refrigeration cycle}

We propose to design a magnetic Otto-type refrigerator with vortices in a type-II superconductor acting as the working substance. Vortices circulate between two heat reservoirs with temperatures~$T_\mathrm H>T_\mathrm C$, extracting heat from the cold reservoir~$\mathrm C$ (on the right in Fig.~\ref{geometry}), and using the hot one~$\mathrm H$ (on the left in Fig.~\ref{geometry}) as a sink. The refrigerator then operates on a four-stroke cycle:
\begin{enumerate}[label=(\arabic*),align=left]
    \item Magnetization: vortices at temperature~$T_\mathrm H$ move from $\mathrm H$ to $\mathrm C$ along the upper arm of the racetrack through a positive magnetic field gradient. This has the effect of cooling down the working substance up to a temperature $T_\mathrm R<T_\mathrm C$.
    \item Heat extraction: the working substance at temperature~$T_\mathrm R$ is put in contact with the cold reservoir~$\mathrm C$. Heat is extracted from the latter as the temperature of the working substance increases until it reaches the reservoir's value~$T_\mathrm C$. The magnetic field is kept at the constant value~$H_\mathrm R$ throughout this step.
    \item Demagnetization: vortices at temperature~$T_\mathrm C$ move along the lower arm of the racetrack through a negative magnetic field gradient. The temperature of the working substance thus rises up to a value~$T_\mathrm L>T_\mathrm H$.
    \item Heat rejection: the working substance at temperature~$T_\mathrm L$ is put in contact with the hot reservoir~$\mathrm H$, in which it rejects heat until its temperature drops to $T_\mathrm H$. The magnetic field is kept at the constant value~$H_\mathrm L$ throughout this step.
\end{enumerate}
In what follows, we assume the temperatures under consideration are much lower than the critical temperature of the superconductor. Additionally, the magnetic field is between the two critical values $H_{c1}<H<H_{c2}$, and close to $H_{c1}$ so the inter-vortex interaction is negligible.  We also assume that the magnetic field gradient and the speed at which vortices move across the racetrack are small enough such that magnetization and demagnetization strokes~(1) and~(3) are performed adiabatically.
% In steady state, the damping coefficient $\eta$ results in an opposing force $f = - \eta v$, where $v$ is the velocity of the fluxon.  The driving force from the current is given by 
% \be
% F = {\vec J} \times \phi_0 {\hat u}
% \ee
% where the unit vector $\hat u$ is parallel to the vortex, and the electrical current $\vec J$ is arranged to flow from the inner edge to the outer edge of the sample.  The magnetic flux quantum is $\phi_0$.  In the steady state, the dissipatative force cancels the Lorentz-like force, resulting in the steady-state velocity of $v = J \phi_0/\eta$.  Thus the power dissipated in the process is given by $f v = J^2 \phi_0^2/\eta$.

% Magnetic work is done on the fluxon as it circulated because the magnetic field is changing.  The number of fluxons is determined by the magnetic flux, $\Phi = BA$. Assisgning each fluon one flux quantum, the number of fluxons in a region is given by $N = BA /\phi_0$, growing linearly with magnetic field.  By increasing the field locally, we thus increase the density of fluxons.  The work required to change the value of magnetic field is given by
% \be
% W  =\int B dM = \mu_0 \int B dN = \frac{\mu_0^2 A}{2  \phi_0}(H_f^2 - H_i^2).
% \ee
% Here the work is done when the fluxon travels along paths of spatially changing magnetic field.

\subsection{$s$-wave model}

We first consider the case of a conventional $s$-wave superconductor. A thermodynamic analysis of the engine cycle is possible given knowledge of the equation of state of the fluxons. In this situation, the main contribution to the specific heat comes from the vortex cores, each of them contributing a constant to the total specific heat which is proportional to the vortex density as a result~\cite{wen_specific_2020}. We then write the specific heat\footnote{Hereafter, all thermodynamic quantities are considered per unit volume.} as:
\begin{equation}
    C(T,H)=\gamma HT,
\end{equation}
where $\gamma$ is a constant. Given the definition $C(T,H)=T\partial S/\partial T|_H$, we can take the entropy to be
\begin{equation}
    S(T,H)=\gamma HT.
    \label{entropy_s}
\end{equation}
The internal energy of the superconductor is naturally expressed in terms of the entropy~$S$ and the magnetization~$M$ as its differential reads
\begin{equation}
    \mathrm dU=T\,\mathrm dS+\mu_0H\,\mathrm dM,
    \label{dU-SM}
\end{equation}
where $\mu_0$ denotes the vacuum permeability. However, the magnetization is not a variable adapted to the situation at stake here. Instead, the relevant variables for our analysis are the entropy~$S$ and the magnetic field~$H$ since strokes~(1) and~(3) take place at constant entropy and strokes~(2) and~(4) take place at constant magnetic field. The energy differential is Eq.~\eqref{dU-SM} above is then rewritten as
\begin{equation}
    \mathrm dU=\left(T+\mu_0H\left.\frac{\partial M}{\partial S}\right|_H\right)\mathrm dS+\mu_0H\left.\frac{\partial M}{\partial H}\right|_S\mathrm dH.
\label{dU-SH}
\end{equation}
The partial derivative of the magnetization with respect to the entropy can be obtained using the Maxwell relation for the magnetic enthalpy~$K=U-\mu_0HM$. Indeed, the differential of $K$ reads
\begin{equation}
    \mathrm dK=T\,\mathrm dS-\mu_0M\,\mathrm dH.
\end{equation}
Invoking the symmetry of second derivatives, we then find
\begin{equation}
    \left.\frac{\partial M}{\partial S}\right|_H=-\frac1{\mu_0}\left.\frac{\partial T}{\partial H}\right|_S=\frac S{\mu_0\gamma H^2},
    \label{dMdS}
\end{equation}
where we have used the explicit expression for the entropy from Eq.~\eqref{entropy_s}. We deduce that the magnetization can be written as
\begin{equation}
    M(S,H)=\frac{S^2}{2\mu_0\gamma H^2}+A(H),
\end{equation}
where $A(H)$ represents the integration constant for the integration with respect to $S$ and is thus a function of $H$. It is not possible to find an explicit expression for $A(H)$ using Maxwell relations because $M$ and $H$ are conjugate variables; it can only be obtained using a microscopic model. We can finally rewrite the energy differential in Eq.~\eqref{dU-SH} as follows:
\begin{equation}
    \mathrm dU=\frac{2S}{\gamma H}\,\mathrm dS-\left(\frac{S^2}{\gamma H^2}-\mu_0H\frac{\mathrm dA}{\mathrm dH}\right)\mathrm dH.
\end{equation}

The energy changes during strokes~(1) and~(3) correspond to the magnetic work done on the superconductor in order to fuel the refrigeration process. We assume that these magnetization and demagnetization steps are performed adiabatically. Hence, $S=S_1=\gamma H_\mathrm LT_\mathrm H$ throughout stroke~(1) and $S=S_3=\gamma H_\mathrm RT_\mathrm C$ throughout stroke~(3). We then derive the magnetic work:
\begin{equation}
\begin{aligned}[t]
    W&=\int_{(1)}\left.\mathrm dU\right|_{S=S_1}+\int_{(3)}\left.\mathrm dU\right|_{S=S_3}\\
    &=-\frac{S_1^2-S_3^2}{\gamma}\int_{H_\mathrm L}^{H_\mathrm R}\frac{\mathrm dH}{H^2}\\
    &=\gamma(H_\mathrm R-H_\mathrm L)\left(\frac{H_\mathrm R}{H_\mathrm L}T_\mathrm C^2-\frac{H_\mathrm L}{H_\mathrm R}T_\mathrm H^2\right).
\end{aligned}
\label{W-sw}
\end{equation}

Further, the temperature of the working substance throughout strokes~(1) and~(3) can be easily calculated using the fact that the entropy remains constant. In particular, the temperatures~$T_\mathrm R$ and~$T_\mathrm L$ at the end of strokes~(1) and~(3) respectively are given by
\begin{align}
     &T_\mathrm R=\frac{S_1}{\gamma H_\mathrm R}=\frac{H_\mathrm L}{H_\mathrm R}T_\mathrm H,\label{swave_TR}\\
     &T_\mathrm L=\frac{S_3}{\gamma H_\mathrm L}=\frac{H_\mathrm R}{H_\mathrm L}T_\mathrm C.\label{swave_TL}
\end{align}
%For refrigeration, we need $T_\mathrm R<T_\mathrm C$ (or $T_\mathrm L>T_\mathrm H$). Therefore, we must have $H_\mathrm L/H_\mathrm R<T_\mathrm C/T_\mathrm R$.

During stroke~(2), the magnetic field stays constant: $H=H_\mathrm R$, while the entropy varies from $S_1$ to $S_3$. The energy variation during this step corresponds to the heat extracted from the cold reservoir:
\begin{equation}
\begin{aligned}[t]
    Q_\mathrm C=\int_{(2)}\left.\mathrm dU\right|_{H=H_\mathrm R}&=\frac2{\gamma H_\mathrm R}\int_{S_1}^{S_3}S\,\mathrm dS\\
    &=\gamma H_\mathrm R\left(T_\mathrm C^2-\frac{H_\mathrm L^2}{H_\mathrm R^2}T_\mathrm H^2\right).
    \end{aligned}
    \label{sw_qc}
\end{equation}
Refrigeration takes place when heat is extracted from the cold reservoir, that is $Q_\mathrm C\ge0$. We see in Eq.~\eqref{sw_qc} above that this imposes
\begin{equation}
    \frac{H_\mathrm L}{H_\mathrm R}\le\frac{T_\mathrm C}{T_\mathrm H}.
    \label{Ineq_HT}
\end{equation}
We also note that $W\ge0$ when the condition in Eq.~\eqref{Ineq_HT} above is satisfied.

The heat rejected in the hot reservoir during stroke~(4) can be calculated in a similar manner: During this step, the magnetic field is constant during stroke~(4): $H=H_\mathrm L$, while the entropy goes from $S_3$ to $S_1$. We then find that
\begin{equation}
    Q_\mathrm H=\int_{(4)}\left.\mathrm dU\right|_{H=H_\mathrm L}=\gamma H_\mathrm L\left(T_\mathrm H^2-\frac{H_\mathrm R^2}{H_\mathrm L^2}T_\mathrm C^2\right).
    \label{sw_qh}
\end{equation}
We immediately see that $Q_\mathrm H<0$ when the condition in Eq.~\eqref{Ineq_HT} is satisfied. Further, it is straightforward to check that $W+Q_\mathrm C+Q_\mathrm H=0$ which corresponds to the first law of thermodynamics: The variation of internal energy during a cycle is zero. As for the second law of thermodynamics, it can be expressed via the Clausius inequality:\footnote{The right-hand side of Eq.~\eqref{2ndLaw} is zero because it corresponds to the variation of entropy during a cycle.}
\begin{equation}
    \frac{Q_\mathrm C}{T_\mathrm C}+\frac{Q_\mathrm H}{T_\mathrm H}\le0.
    \label{2ndLaw}
\end{equation}
One can check that the expressions for $Q_\mathrm C$ and $Q_\mathrm H$ in Eqs.~\eqref{sw_qc} and~\eqref{sw_qh} respectively satisfy the inequality in Eq.~\eqref{2ndLaw} above.

The refrigerator's coefficient of performance ($COP$) is given by the ratio of the heat extracted from the cold reservoir to the work supplied to the superconductor:
\begin{equation}
    COP=\frac{Q_\mathrm C}W=\frac{H_\mathrm L}{H_\mathrm R-H_\mathrm L}=\frac{\mu}{1-\mu},
    \label{cop_sw}
\end{equation}
 similar to Otto cycle results, where $\mu=H_\mathrm L/H_\mathrm R$. Using the first and second laws of thermodynamics, one can show that the refrigerator's $COP$ is upper bounded by the Carnot $COP$ which is reached when the refrigerator operates reversibly:
\begin{equation}
    COP=\frac{\mu}{1-\mu}\le COP_\mathrm{Carnot}=\frac{T_\mathrm C}{T_\mathrm H-T_\mathrm C}.
    \label{boundCOP}
\end{equation}
Carnot's theorem in Eq.~\eqref{boundCOP} is equivalent to Eq.~\eqref{Ineq_HT} in the situation at stake. We then find that Carnot $COP$ is reached when $\mu=T_\mathrm C/T_\mathrm H$, in which case $Q_\mathrm C=0$.

\subsection{$d$-wave Model}

In the case of a $d$-wave superconductor, the specific heat calculated using the Volovik model is given by~\cite{wen_specific_2020,volovik93}
\be
C(T,H)=\alpha\sqrt HT, 
\ee
where $\alpha$ is again a constant. Applying the same treatment as in the case of an $s$-wave superconductor, we first calculate the entropy:

\be
S=\alpha\sqrt HT.
\label{S_expr}
\ee
%Therefore, adiabatic transitions are given by constant entropy, so we conclude the first leg of the journey (moving from left to right along the upper path), starting at a magnetic field $H_L$ and transitioning to a larger magnetic field $H_R$ will result in the temperature of the fluxon dropping to 
% \be
% T_R = T_H \sqrt{H_L /H_R}.
% \label{Tr} 
% \ee
% We demand that the gradient in magnetic field is strong enough so that $H_R > H_L (T_H/T_C)^2$.
% This is the cooling power stroke of the cycle that cools the fluxons below the temperature of the cold bath which it will refrigerate.
We can now obtain a formal expression for the magnetization. We have
\begin{equation}
    \left.\frac{\partial M}{\partial S}\right|_H=-\frac1{\mu_0}\left.\frac{\partial T}{\partial H}\right|_S=\frac S{2\mu_0\alpha H^{3/2}},
\end{equation}
which leads to
\be
    M(S,H)=\frac{S^2}{4\mu_0\alpha H^{3/2}}+A(H),
\ee
where $A(H)$ is an unknown function of the magnetic field. As a consequence, the differential for the internal energy expressed in the relevant variables~$S$ and~$H$ is given by
\begin{equation}
    \mathrm dU=\frac{3S}{2\alpha\sqrt H}\,\mathrm dS-\left(\frac{3S^2}{8\alpha H^{3/2}}-\mu_0H\frac{\mathrm dA}{\mathrm dH}\right)\mathrm dH.
\end{equation}

The temperatures at the end of strokes~(1) and~(3) can be again obtained using the constant entropy condition:
\begin{align}
    &T_\mathrm R=\sqrt{\frac{H_\mathrm L}{H_\mathrm R}}T_\mathrm H,\\
    &T_\mathrm R=\sqrt{\frac{H_\mathrm R}{H_\mathrm L}}T_\mathrm C.
\end{align}

The work supplied to the superconductor during a cycle is then calculated as in Eq.~\eqref{W-sw}, and we obtain
\begin{equation}
    W=\frac{3\alpha}4\left(\sqrt{H_\mathrm R}-\sqrt{H_\mathrm L}\right)\left(\sqrt\frac{H_\mathrm R}{H_\mathrm L}T_\mathrm C^2-\sqrt\frac{H_\mathrm L}{H_\mathrm R}T_\mathrm H^2\right).
\end{equation}
Conversely, strokes~(2) and~(4) take place at constant magnetic fields, and we calculate $Q_\mathrm C$ and $Q_\mathrm H$ as in Eqs.~\eqref{sw_qc} and~\eqref{sw_qh} respectively to find
\begin{align}
    &Q_\mathrm C=\frac{3\alpha\sqrt{H_\mathrm R}}4\left(T_\mathrm C^2-\frac{H_\mathrm L}{H_\mathrm R}T_\mathrm H^2\right),\label{Qc}\\
    &Q_\mathrm H=\frac{3\alpha\sqrt{H_\mathrm L}}4\left(T_\mathrm H^2-\frac{H_\mathrm R}{H_\mathrm L}T_\mathrm C^2\right).
\end{align}
It is clear from Eq.~\eqref{Qc} above that refrigeration is only possible when
\begin{equation}
    \frac{H_\mathrm L}{H_\mathrm R}\le\frac{T_\mathrm C^2}{T_\mathrm H^2},
    \label{eqineq}
\end{equation}
in which case $Q_\mathrm C\ge0$, $Q_\mathrm H\le0$ and $W\ge0$.

Finally, the refrigerator's $COP$ is given by
\begin{equation}
    COP=\frac{Q_\mathrm C}W=\frac{\sqrt{H_\mathrm L}}{\sqrt{H_\mathrm R}-\sqrt{H_\mathrm L}}=\frac{\sqrt\mu}{1-\sqrt\mu}.
    \label{cop_dw}
\end{equation}
Similarly to the case of an $s$-wave superconductor, the refrigeration condition in Eq.~\eqref{eqineq} is equivalent to $COP\le COP_\mathrm{Carnot}$. The Carnot coefficient of performance~$COP_\mathrm{Carnot}$ is achieved when equality is realized in Eq.~\eqref{eqineq}, in which case $Q_\mathrm C=0$.\par

\begin{figure}
    \centering
    \includegraphics[width=\linewidth]{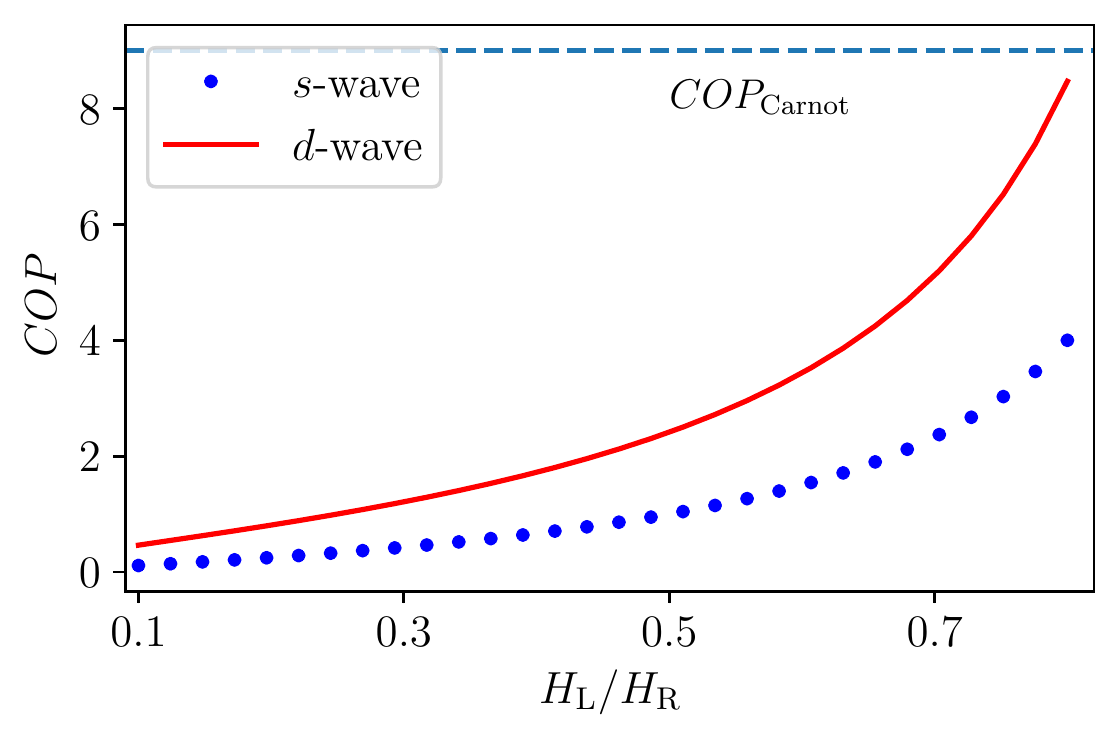}
    \caption{Plot of the fluxon refrigerator $COP$ as a function of the ratio of the magnetic fields on the left and right arms. The blue dots (red line) shows the $COP$ for an $s$-wave ($d$-wave) superconductor. We see that as the magnetic fields come closer to one another, the $COP$ increases. Also, the $COP$ for a $d$-wave superconductor is higher than that of an $s$-wave superconductor under the same conditions. For both $s$ and $d$-wave cases, the reservoir temperatures set the maximum possible value of the magnetic field ratio (see Eqs.~\eqref{Ineq_HT} and \eqref{eqineq}).  The dashed horizontal line corresponds to $COP_\mathrm {Carnot}$ when $T_\mathrm C/T_\mathrm H=0.9$.}
    \label{cop_vs_mu0}
\end{figure}

\subsection{Results}

From Eqs.~\eqref{cop_sw} and~\eqref{cop_dw}, it is clear that the coefficient of performance only depends on the applied magnetic fields for both $s$-wave and $d$-wave superconductors. Fig.~\ref{cop_vs_mu0} shows the behavior of the coefficient of performance for both of these cases as functions of the magnetic field ratio~$H_\mathrm L/H_\mathrm R$. The $COP$ clearly increases with this ratio and reaches its maximal value when the refrigerator operates reversibly, in which case no refrigeration occurs. This is achieved when $H_\mathrm L/H_\mathrm R=T_\mathrm C/T_\mathrm H$ for $s$-wave superconductors and $H_\mathrm L/H_\mathrm R=T_\mathrm C^2/T_\mathrm H^2$ for $d$-wave superconductors. Additionally, both kinds of superconductors exhibit similar behaviors, although $d$-wave superconductors show slightly better performance. For simplicity,  we focus on $s$-wave superconductors throughout the rest of our analysis. Analogous explorations can be considered for $d$-wave superconductors and are expected to lead to similar qualitative conclusions.

\section{\label{dynamics}Fluxon propagation and dissipation}
In reality, this simple model above needs to be augmented to take into account the energy loss of the fluxons as they are being driven around the racetrack, and the effects of vortex thermalization at finite velocities.

\subsection{Propagation of fluxons in the steady state}

Taking a mesoscopic sample of the fluxon lattice as the working substance, we first consider the following simple form for the heat capacity per unit volume for $s$-wave superconductors  $C(T,{ H}) = \gamma HT$. The entropy is $S(T,{ H})=\gamma HT$.

Given that entropy $S(T,H)$ is a state function, it can be written as an exact differential of the form, 
\begin{equation}
\frac{\mathrm dS}{\mathrm dx} =\frac{\partial S}{\partial T}\frac{\mathrm dT}{\mathrm dx} + \frac{\partial S}{\partial { H}}\frac{\mathrm dH}{\mathrm dx}=\gamma ({ H}\partial_x T+T\partial_x{ H}), 
\end{equation} 
where $x$ is the length along the arms.% For an adiabatic process, we have $\frac{dS}{dx} = 0$. This immediately predicts the temperature drop per spatial variation of the magnetic field, for any applied magnetic field profile, as,
%\begin{eqnarray}
%    &\partial_{x}T(x)=-\frac{ T(x)\partial_{x}{ H}}{ { H}}\\&\implies T(x)=T(0)\frac{{ H}(x)}{{ H}(0)},
%\end{eqnarray}
%which is consistent with Eqs.~\eqref{swave_TR} and~\eqref{swave_TL}.
\par Recall that adiabatic processes are processes where the system is supposed to be thermally isolated from the environment. 
Heat losses can be taken to account by modifying  $\frac{dS}{dx} = 0$ to  $\frac{dS}{dx} = \tfrac{1}{v}\frac{P(x)}{T(x)}$, where $P(x)$ is the power loss due to dissipation along an arm and $v$ is the speed of fluxons (also see App.~\ref{appA}).  The differential equations modify to,
\begin{eqnarray}
    \frac{P(x)}{T(x)} &=& \gamma v ({ H}(x)\partial_x T(x)+T(x)\partial_x{H}(x)).
    \label{power_dyn}
\end{eqnarray}
%Rearranging we get,
%\begin{equation}
 %   C(T(x),{ H}(x)) \partial_{x}T(x)= \left(\frac{1}{v}P(x) - \gamma T^{2}(x)\partial_{x}{ H}(x)\right).
%    \label{c_dynamic_s}
%\end{equation}
This can be solved for different adiabatic and isothermal arms of the cycle, accounting for additional dissipation mechanisms in $P(x)$. \par 
 %Similar calculations, for $d$-wave superconductors lead to
% \begin{equation}
  %  C(T,{ H}) \partial_{x}T = \frac{1}{v}P(x) - \frac{\alpha T^{2}(x)}{2\sqrt{{H}(x)}}\partial_{x}{H}(x).
  %  \label{c_dynamic_d}
%\end{equation}

\subsection{\label{dissipation}Incorporating dissipation}
 In steady state, the damping coefficient $\eta$ results in an opposing force $f = - \eta v$.  The driving force from the current is given by 
 \be
 F = {\vec J} \times \phi_0 {\hat u},
 \ee
 where the unit vector $\hat u$ is parallel to the vortex, and the electrical current $\vec J$ is arranged to flow from the inner edge to the outer edge of the sample.  $\phi_0=\tfrac{h}{2e}$  is the magnetic flux quantum.  In the steady state, the dissipative force cancels the Lorentz-like force, resulting in the steady-state velocity of $v = J \phi_0/\eta$. % Thus the power dissipated in the process is given by $f v = J^2 \phi_0^2/\eta$.
We can utilize the discussion in the previous section to incorporate the effect of dissipation in our description. In this case $P(x)=n(x)\eta v^2$, where $n(x)$ is the surface density of fluxons at position $x$  \cite{manginkahn, alma991011414589703276}. Therefore, for step (1) we have (in the steady state), 

\begin{equation}
    \frac{\partial S}{\partial x}=\frac{n(x)\eta v}{ T(x)}.
    \label{S_eqn}
\end{equation}
Next, we assume  $n(x)=n_\mathrm L\left(1+\tfrac{x}{L_x}(1/\mu-1)\right)$, where $n_L$ is the fluxon density in the arm in contact with the hot reservoir and $L_x$ is the length of the arms along which there is a magnetic field gradient (i.e.~arms (1) and (3)). In other words, assuming linearly varying magnetic field, and using expression \eqref{entropy_s}, for $s$-wave superconductors we get
\begin{equation}
    T_\mathrm R=\left(\mu^2\tilde{T}_H^2 + \frac{2\mu_0 J  L_x}{3\gamma}(1+\mu+\mu^2)\right)^{1/2}.
    \label{Tr_diss}
\end{equation}
Here $\tilde{T}_\mathrm H$ is the final temperature of the fluxons after step~(4), and could be different from $T_\mathrm H$ due to dissipation (as we will see soon).
Similarly, after step~(3)
\begin{equation}
    T_\mathrm L=\left(\frac{\tilde{T}_C^2}{\mu^2} + \frac{2\mu_0 J  L_x}{3\gamma\mu^2}(1+\mu+\mu^2)\right)^{1/2},
    \label{Tl_diss}
\end{equation}
where $\tilde{T}_\mathrm C$ is the final temperature of fluxons after stroke (2). Without dissipation, Eqs.~\eqref{Tr_diss} and~\eqref{Tl_diss} reduce to Eqs.~\eqref{swave_TR} and~\eqref{swave_TL}, as expected.

\subsection{Simple dynamical model for heat exchanges}

In this section, we introduce a simple model to analyze the dynamics of the heat exchanges between the reservoirs and the working substance as the latter moves along the superconductor geometry. %We consider an infinitesimal slice of length~$\mathrm dy$ in the direction of the racetrack.
At the beginning of stroke~(2), $y=0$, its temperature is $T_\mathrm R$, and it is coupled to the cold reservoir at temperature~$T_\mathrm C>T_\mathrm R$.

The working substance moves at a constant speed~$v$ and spends a time~$L_y/v$ in contact with the reservoir, where $L_y$ is the length of the racetrack in the relevant direction. The magnetic field does not change during this process: $H=H_\mathrm R$. As the working substance moves along the track, it receives heat from the reservoir and dissipates energy due to vortex drag. In time $\mathrm dt$, the working substance moves a distance $\mathrm dy=v\mathrm dt$ along the track. During this infinitesimal distance travelled, both the heat exchange with the reservoir and the heat generated due to electrical current contributes to the change in the fluxons' internal energy. The energy balance (per unit volume) in interval~$\mathrm dy$ along the racetrack reads
\begin{equation}
    U(y+\mathrm dy)-U(y)=\left(\dot Q+n_\mathrm R\eta v^2\right)\frac{\mathrm dy}v,
    \label{energybalance}
\end{equation}
where $U(y)$ is the energy density of the working substance at position $y$, $\dot Q$ is the heat current per unit volume from the reservoir, and $n_\mathrm R$ is the vortex density on the right arm. The second term to the right-hand side above corresponds to the energy transferred to the working substance so that the vortex speed~$v$ remains constant despite the drag force. This is done practically by imposing an electrical current flow in the direction perpendicular to the vortex motion.

We introduce a phenomenological Fourier-type formula for the heat current~$\dot Q$:
\begin{equation}
    \dot Q=-\kappa_\mathrm{th}(T-T_\mathrm C),
    \label{Fourier}
\end{equation}
where $T$ is the temperature of the working substance, and $\kappa_\mathrm{th}$ denotes the thermal conductivity (per unit volume) at the contact with the reservoir. The thermal conductivity is assumed to be temperature-independent for simplicity. The expression for the heat current in Eq.~\eqref{Fourier} above is naturally valid when the temperature difference~$T-T_\mathrm C$ is small, but it often holds outside this regime.

Finally, we choose temperature and magnetic field as the relevant variables to describe the state of the working substance. As a consequence, the energy density is expressed through these quantities:
\begin{equation}
    U(y)=U(T(y),H(y))=U(T(y),H_\mathrm R).
\end{equation}
We can now rewrite the energy time derivative as follows:
\begin{equation}
\begin{aligned}[t]
    \frac{\mathrm dU}{\mathrm dy}=\frac{\mathrm dT}{\mathrm dy}\left.\frac{\partial U}{\partial T}\right|_H&=\left(T\left.\frac{\partial S}{\partial T}\right|_H+\mu_0H\left.\frac{\partial M}{\partial T}\right|_H\right)\frac{\mathrm dT}{\mathrm dy}\\
    &=2\gamma HT\frac{\mathrm dT}{\mathrm dy},
    \end{aligned}
    \label{dudt}
\end{equation}
where $\partial M/\partial T|_H$ has been calculated using the same technique that yielded Eq.~\eqref{dMdS}. The energy balance in Eq.~\eqref{energybalance} then becomes
\begin{equation}
    2v\gamma H_\mathrm RT\frac{\mathrm dT}{\mathrm dy}=-\kappa_\mathrm{th}(T-T_\mathrm C)+n_\mathrm R\eta v^2.
\end{equation}
As a consequence, the temperature of the working substance obeys the following differential equation
\begin{equation}
    \frac{\mathrm dT}{\mathrm dy}=\frac{\kappa_\mathrm{th} T_\mathrm C+n_\mathrm R\eta v^2}{2v\gamma H_\mathrm RT}-\frac{\kappa_\mathrm{th}}{2v\gamma H_\mathrm R}.
    \label{dTdt_R}
\end{equation}
The differential equation above can be solved analytically, and we find that the temperature reads
\begin{equation}
    T(y)= T_{\mathrm f\mathrm R}\left(1+W\left((\tfrac{T_\mathrm R}{T_{\mathrm f\mathrm R}}-1)\e^{\tfrac{T_\mathrm R}{T_{\mathrm f\mathrm R}}-1-y/v\tau_\mathrm R}\right)\right),
    \label{Tt_R}
\end{equation}
where $W$ denotes the Lambert function, while the constants~$\tilde T_\mathrm{fR}$ and~$\tau_\mathrm R$ are given by
\begin{align}
    & T_{\mathrm f\mathrm R}=T_\mathrm C+\frac{n_\mathrm R\eta v^2}{\kappa_\mathrm{th}},\quad\tau_\mathrm R=\frac{2\gamma H_\mathrm R T_{\mathrm f\mathrm R}}{\kappa_\mathrm{th}}.
\end{align}
$ T_{\mathrm f\mathrm R}$ is the final temperature reached by the working substance if it were to stay in contact with reservoir~$\mathrm C$ for an infinitely long time, and $\tau_\mathrm R$ is the characteristic time over which the working substance reaches its final temperature.

We observe that the asymptotic final temperature of the working substance~$T_{\mathrm f\mathrm R}$ is higher than $T_\mathrm C$ when dissipation is taken into account. As a result, the heat current~$\dot Q$ starts to flow from the working substance to the reservoir after some time to the detriment of refrigeration. When the working substance travels slowly across the racetrack, the final temperature approaches the reservoir temperature:
\begin{equation}
   T_{\mathrm f\mathrm R}\simeq T_\mathrm C\quad\text{if}\quad v\ll\sqrt{\frac{\kappa_\mathrm{th} T_\mathrm C}{n_\mathrm R\eta}}.
    \label{Tf=TC}
\end{equation}

The total heat exchanged with the cold reservoir is given by
\begin{equation}
    Q_\mathrm C=\int_0^{L_y}\mathrm dy\,\dot Q/v=-\tfrac{\kappa_\mathrm{th}}{v}\int_0^{L_y}\mathrm dy\,(T(y)-T_\mathrm C).
\end{equation}
The integration can be carried out analytically, and we find
\begin{equation}
 \begin{split}
     Q_\mathrm C&=\frac{\kappa_\mathrm{th} L_y}v(T_\mathrm C-T_{\mathrm f\mathrm R})-\frac{\kappa_\mathrm{th}\tau_\mathrm R T_{\mathrm f\mathrm R}}2\left(\tfrac{T_\mathrm R}{T_{\mathrm f\mathrm R}}\right)^2\\&+\frac{\kappa_\mathrm{th}\tau_\mathrm R T_{\mathrm f\mathrm R}}2\left(1+W\left((\tfrac{T_\mathrm R}{T_{\mathrm f\mathrm R}}-1)\e^{\tfrac{T_\mathrm R}{T_{\mathrm f\mathrm R}}-1-L_y/v\tau_\mathrm R}\right)\right)^2.
   \end{split}
   \label{QC_dyn}
\end{equation}
In a  similar manner, stroke~(4) leads to cooling of the working substance from $T_\mathrm L$ to a final temperature $T_\mathrm{fL}$. In this case, Eq.~\eqref{dudt} takes the form  (assuming the same thermal conductivity)
\begin{equation}
\begin{split}
    \frac{\mathrm dU}{\mathrm dy}&=\frac{\mathrm dT}{\mathrm dy}\left.\frac{\partial U}{\partial T}\right|_H\\\implies 2v\gamma H_\mathrm L T\frac{\mathrm dT}{\mathrm dy}&=-\kappa_\mathrm{th}(T-T_\mathrm H)+n_\mathrm L\eta v^2.
\end{split}
    \label{dudt1}
\end{equation}
Therefore, 
\begin{equation}
    T(y)=T_{\mathrm f \mathrm L}\left(1+W\left((\tfrac{T_\mathrm L}{T_{\mathrm f \mathrm L}}-1)\e^{\tfrac{T_\mathrm L}{T_{\mathrm f \mathrm L}}-1-y/v\tau_\mathrm L}\right)\right),
    \label{Tt_L}
\end{equation}
where the constants~$T_{\mathrm f \mathrm L}$ and~$\tau_\mathrm L$ are given by
\begin{align}
    &T_{\mathrm f \mathrm L}=T_\mathrm H+\frac{n_\mathrm L\eta v^2}{\kappa_\mathrm{th}},\quad\tau_\mathrm L=\frac{2\gamma { H}_\mathrm LT_{\mathrm f\mathrm L}}{\kappa_\mathrm{th}}.
    \label{const_L}
\end{align}
Heat transferred to the hot reservoir is given by
\begin{equation}
\begin{aligned}[t]
 & Q_\mathrm H=
   \frac{\kappa_\mathrm{th} L_y}v(T_{\mathrm f\mathrm L}-T_\mathrm H)+ \frac{\kappa_\mathrm{th}\tau_\mathrm L T_{\mathrm f\mathrm L}}2\left(\tfrac{T_\mathrm L}{T_{\mathrm f \mathrm L}}\right)^2\\
   &-\frac{\kappa_\mathrm{th}\tau_\mathrm L T_{\mathrm f\mathrm L}}2\left(1+W\left((\tfrac{T_\mathrm L}{T_{\mathrm f \mathrm L}}-1)\e^{\theta_\mathrm L-1-L_y/v\tau_\mathrm L}\right)\right)^2.
   \end{aligned}
   \label{QH_dynamic}
\end{equation}

\section{\label{quanti} Quantitative analysis of the heat exchanged and refrigerator performance}
With all the necessary physics at hand, we now analyze how the fluxon dynamics influence the heat exchanged and the refrigerator's performance.  
\subsection{\label{dimensionless} Refrigeration cycle description in terms of dimensionless control parameters}
We can analyze the system's behavior by looking at five independent dimensionless parameters
\begin{equation}
\begin{split}
       & l=L_y/L_x,\quad \mu= { H}_\mathrm L/{ H}_\mathrm R,\quad \phi_\mathrm C=T_\mathrm C/T_\mathrm H \\& \bar{v}=v\frac{2\gamma{ H}_\mathrm R T_\mathrm H}{\kappa_\mathrm{th} L_x}, \quad \Omega=\frac{\mu_0\kappa_\mathrm{th}\eta L_x^2}{2\phi_0 \gamma^2{ H}_{\mathrm R}T_{\mathrm H}^3}.
    \label{params}
\end{split}
\end{equation}
Here $l$ signifies the geometry of the superconducting material. $\mu$ and $\phi_\mathrm C$ are the magnetic field ratio and reservoir temperatures ratio respectively. $\bar{v}$ is the dimensionless speed of the fluxons. $\Omega$ is a parameter signifying dissipation and depends on superconductor properties, length along $x$, magnetic field on the right-hand side, and hot reservoir temperature.  We also define temperatures scaled with respect to the hot reservoir temperature as $\phi=T/T_\mathrm H$. 
The temperatures at the end of stroke~(2) and~(4) are
\begin{equation}
    \begin{split}
        &\tilde{\phi}_\mathrm C=\phi_{\mathrm f\mathrm R}\left(1+W\left(\left(\tfrac{\phi_\mathrm R}{\phi_{\mathrm f\mathrm R}}-1\right)e^{\tfrac{\phi_\mathrm R}{\phi_{\mathrm f\mathrm R}}-1-\tfrac{l}{\bar{v}\phi_{\mathrm f\mathrm R}}}\right)\right),\\
        &\tilde{\phi}_\mathrm H=\phi_{\mathrm f\mathrm L}\left(1+W\left(\left(\tfrac{\phi_\mathrm L}{\phi_{\mathrm f\mathrm L}}-1\right)e^{\tfrac{\phi_\mathrm L}{\phi_{\mathrm f\mathrm L}}-1-\tfrac{l}{\mu\bar{v}\phi_{\mathrm f\mathrm L}}}\right)\right),
    \end{split}
    \label{phit_hc}
\end{equation}
\begin{figure}
    \centering
    \includegraphics[width=\linewidth]{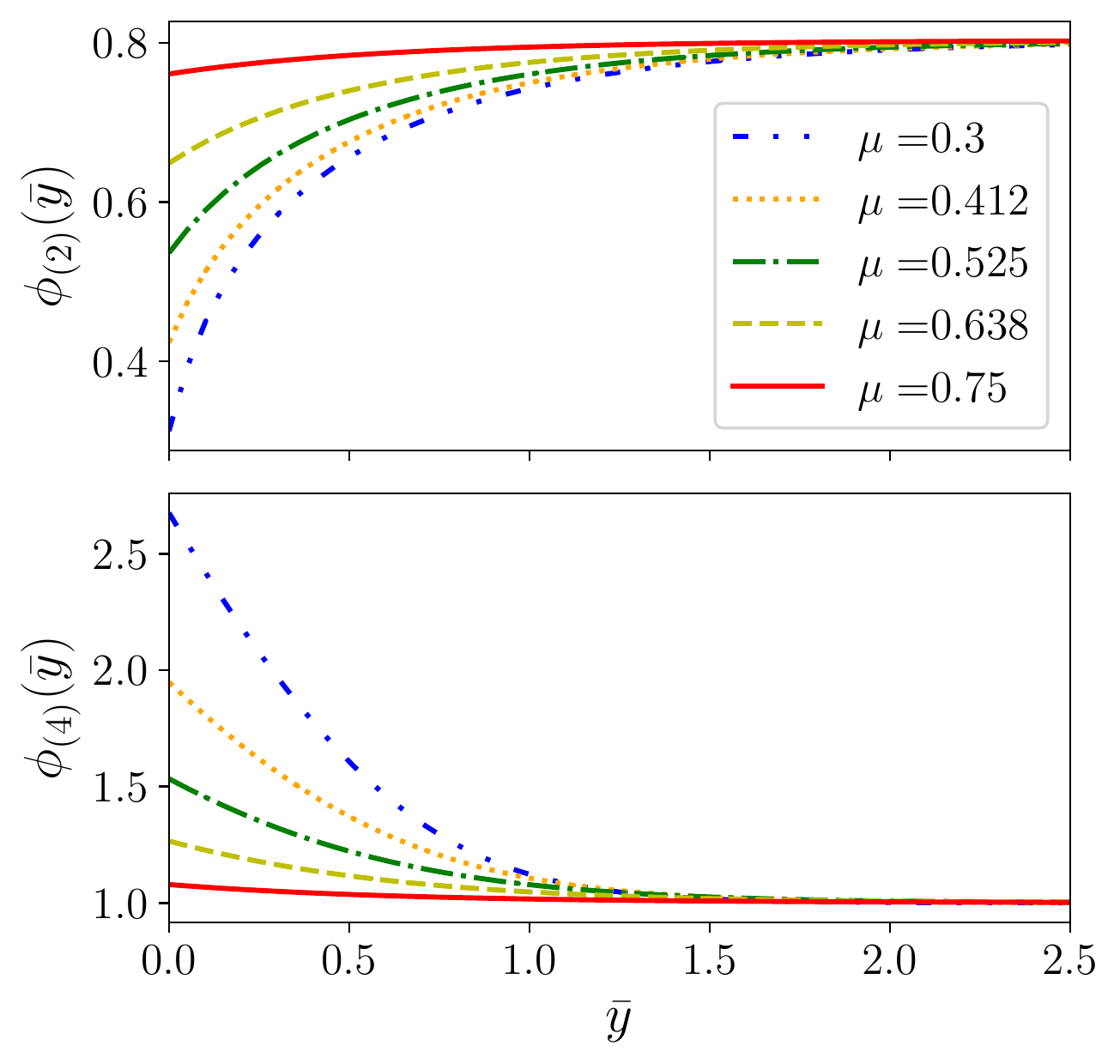}
    \caption{Temperature as a function of scaled length $\bar{y}=y/L_x$ along the arms in contact with reservoirs in stroke 2 (top) and 4 (bottom) for different values of the magnetic field ratio. Here we have chosen $\phi_\mathrm C=0.8$, $l=2.5$, $\bar{v}=0.8$, $\Omega=0.01$. We see that the temperature in stroke 2 (4) increases (decreases) and tends to a value close to $\phi_\mathrm C$ ($\phi_\mathrm H=1$). With the decrease of the magnetic field, the initial  temperature  decreases (increases) in the top (bottom) plot. This behavior can be understood from Eqs.~\eqref{swave_TR} and \eqref{swave_TL}, where the temperature at the end of stroke 1 (3) is proportional (inversely proportional) to the magnetic field ratio.}
    \label{temp_isotherm}
\end{figure}
with $\phi_{\mathrm f\mathrm R}=\phi_\mathrm C+\tfrac{1}{2}\bar{v}^2\Omega$ and $\phi_{\mathrm f\mathrm L}=1+\tfrac{1}{2}\bar{v}^2\mu\Omega$. Also, Eqs.~\eqref{Tr_diss} and \eqref{Tl_diss} become
\begin{equation}
    \begin{split}
        &\phi_\mathrm R^2=\mu^2\tilde{\phi}^2_\mathrm H+\frac{2\bar{v}\Omega}{3}(1+\mu+\mu^2),\\
        &\phi_\mathrm L^2=\frac{\tilde{\phi}^2_\mathrm C}{\mu^2}+\frac{2\bar{v}\Omega}{3\mu^2}(1+\mu+\mu^2).
    \end{split}
    \label{phi_rl}
\end{equation}
Fig.~\ref{temp_isotherm} shows the behavior of the fluxon temperatures in strokes 2 and 4 as functions of the distance traveled. At the end of the stroke, they tend to the temperatures of corresponding reservoirs.  

The heat extracted from the cold reservoir is
\begin{equation}
    Q_{\mathrm C}=\gamma{ H}_{\mathrm R}T_{\mathrm H}^2\left(-l\bar{v}\Omega+(\tilde{\phi}_{\mathrm C}^2-\phi_{\mathrm R}^2)\right).
    \label{qc_simple}
\end{equation}
 %and cooling power

When the speed~$\bar{v}$ is slow enough, the exponential factor inside the Lambert function is negligible as compared with any integer power of $\bar{v}$. This allows us to neglect the Lambert function in Eq.~\eqref{phit_hc} above to obtain $\tilde{\phi}_\mathrm C\simeq\phi_{\mathrm f\mathrm R}$ and

\begin{equation}
    Q_\mathrm C\simeq\gamma { H}_\mathrm R T_\mathrm H^2\left(-l\bar{v}\Omega+(\phi_{\mathrm f\mathrm R}^2-\phi_{\mathrm R}^2)\right)\quad\text{if}\quad \bar{v}\ll\frac{l}{\phi_C}.
    \label{QC_approx}
\end{equation}

Using the approximate expression in Eq.~\eqref{QC_approx} above, one can check that $Q_\mathrm C$ decreases with $\bar{v}$ at slow speeds. This behavior is well-captured by a first-order expansion in $v$, which amounts to neglecting the difference between $\phi_{\mathrm f\mathrm R}$ and $\phi_\mathrm C$ following the criterion in Eq.~\eqref{Tf=TC} (or equivalently $\bar{v}\ll\sqrt{1/\Omega}$). We then have:
\begin{equation}
   Q_\mathrm C\simeq\gamma { H}_\mathrm R T_\mathrm H^2\left(\left(\phi_\mathrm C^2-\mu^2\right)-\Omega\left(l+\tfrac{2}{3}(1+\mu+\mu^2)\right)\bar{v}\right).
    \label{QC_lin}
\end{equation}

This decrease of $Q_\mathrm C$ with $\bar{v}$ indicates that the refrigerator can only operate at slow speeds. Indeed, $Q_\mathrm C$ becomes negative when $\bar{v}$ exceeds the speed limit~$\bar{v}_{\lim,R}$ given by
\begin{equation}
    \bar{v}_{\lim,R}=\frac{\phi_\mathrm C^2-\mu^2}{\Omega\left(l+\tfrac{2}{3}(1+\mu+\mu^2)\right)}.
    \label{vlim}
\end{equation}
Numerical results using the analytical expression for $Q_\mathrm C$ in Eq.~\eqref{qc_simple} confirm that refrigeration is only possible at slow speeds, with $\bar{v}_{\lim,R}$ in Eq.~\eqref{vlim} above being a good estimate for the exact speed limit. Furthermore, the approximate expression for $Q_\mathrm C$ in Eq.~\eqref{QC_lin} shows that the maximum amount of heat extracted from the cold reservoir is reached for $\bar{v}\to0$. We then find
\begin{equation}
    Q_\mathrm C^{\max}=\gamma { H}_\mathrm R T_\mathrm H^2\left(\phi_\mathrm C^2-\mu^2\right),
    \label{qcmax}
\end{equation}
which is the result in Eq.~\eqref{sw_qc}. Thus, for refrigeration to occur, we must have $\phi_\mathrm C\ge\mu$,  see Eq.~\eqref{Ineq_HT}.

Our dynamical model also allows for the calculation of the refrigerator's cooling power. The latter is given by the ratio of the heat extracted from the cold reservoir to the time necessary to perform such extraction:
\begin{equation}
    P_{\mathrm C}=\frac{Q_\mathrm C}{L_y/v}=\frac{\kappa_\mathrm{th} T_{\mathrm H}}{2}\frac{\bar{v}}{l }\left(-l\bar{v}\Omega+(\tilde{\phi}_{\mathrm C}^2-\phi_{\mathrm R}^2)\right).
    \label{pc_simple}
\end{equation}

We can use Eq.~\eqref{QC_lin} to approximate 
\begin{equation}
    P_{\mathrm C}\simeq\frac{\kappa_\mathrm{th} T_{\mathrm H}}{2}\frac{\bar{v}}{l }\left(\left(\phi_\mathrm C^2-\mu^2\right)-\Omega\left(l+\tfrac{2}{3}(1+\mu+\mu^2)\right)\bar{v}\right).
    \label{pc_approx}
\end{equation}
We then find that the cooling power reaches its maximum value for $\bar{v}=\bar{v}_{\lim,R}/2$, and we have
\begin{equation}
    P_\mathrm C^{\max}\simeq\frac{\kappa_\mathrm{th} T_{\mathrm H}}{2}\frac{\left(\phi_\mathrm C^2-\mu^2\right)^2}{4l\Omega\left(l+\tfrac{2}{3}(1+\mu+\mu^2)\right)}.
    \label{pcmax}
\end{equation}

\begin{figure}
    \centering
    \includegraphics[width=\linewidth]{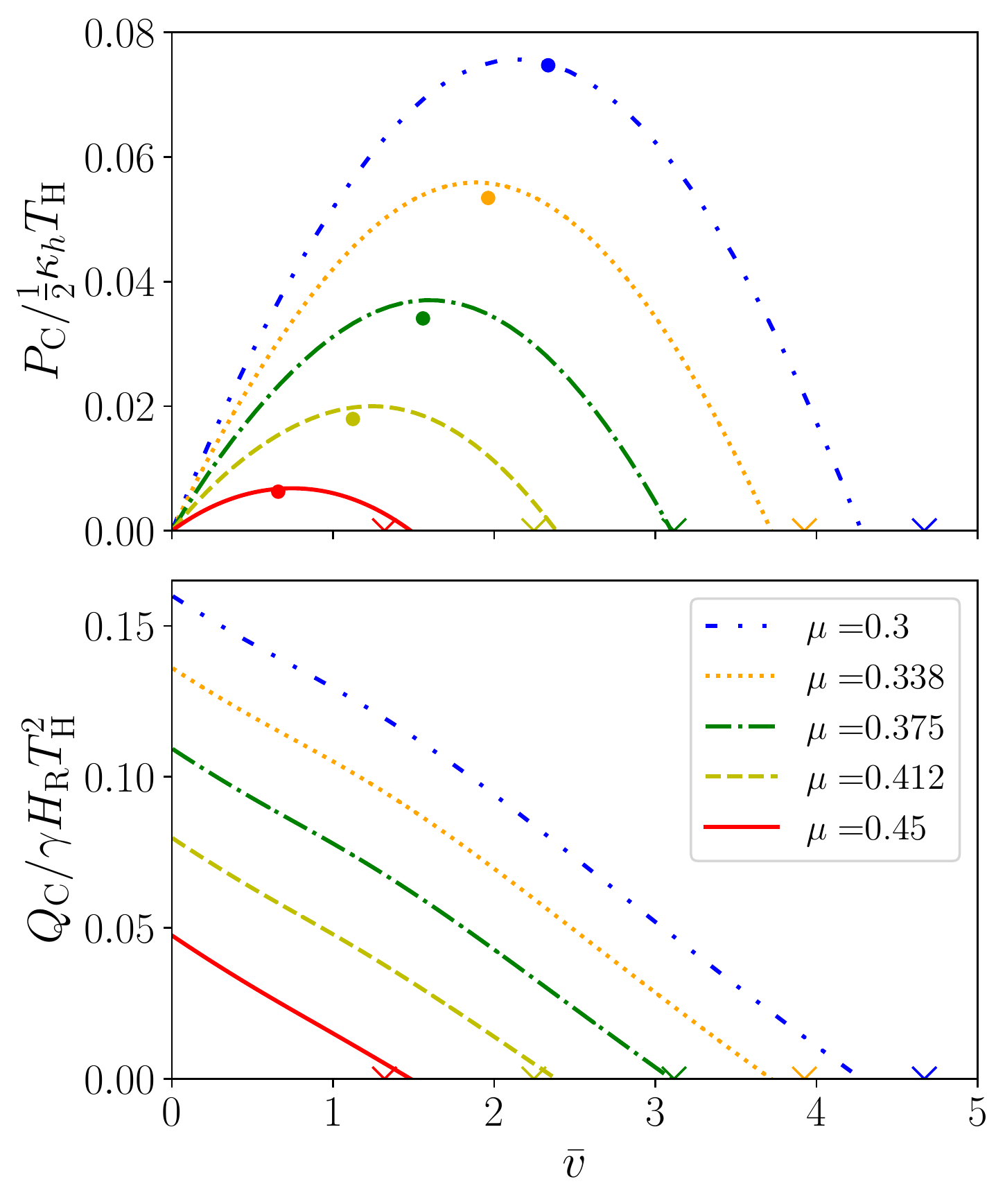}
    	%\begin{tikzpicture}[overlay]
	%\node[] at (-4,11.) {(a)};
	%\node[] at (-4,5.5) {(b)};
	%\end{tikzpicture}
    \caption{Plot of scaled cooling power (top) and heat withdrawn from the cold reservoir (bottom) as  functions of dimensionless fluxon velocity $\bar{v}$ for $l=2.5$, $\Omega=0.01$, $\phi_\mathrm C=0.5$ and different values of $\mu$. The dots in the top panel show the corresponding approximate cooling power maxima calculated analytically in Eq.~\eqref{pcmax}. The crosses show Eq.~\eqref{vlim}, i.e.~the velocity where both $Q_\mathrm C$ and $P_\mathrm C$ are 0. As $\mu$ decreases, the analytical approximations become worse at capturing the behavior. We can understand this by realizing that the $\mathcal{O}(\bar{v}^3)$ term in Eq.~\eqref{pc_simple} has a coefficient proportional to $\phi_\mathrm C-\mu^3$. Therefore, for the chosen parameter regime, as $\mu$ decreases, the higher-order effects become more prominent. Although not shown, the same effect can be observed as we increase the value of $\phi_\mathrm C$.}
    \label{Pcplot}
\end{figure}

It is noteworthy that Eqs.~\eqref{phit_hc} and~\eqref{phi_rl} are implicit in nature, and could be solved numerically for arbitrary parameter values.  In Fig.~\ref{Pcplot}, we show the cooling power as a function of $\bar{v}$ for different values of the magnetic field ratios. 

In Fig.~\ref{Pcplot}, we show the cooling power and the heat withdrawn from the cold reservoir as functions of fluxon speed for different values of the magnetic field ratios. The cooling power has a parabola-like shape as a function of velocity, while the heat exchanged decreases almost linearly. The approximations in Eqs.~\eqref{QC_lin}, \eqref{pc_approx} and~\eqref{pcmax} describe the behavior of the heat withdrawn and cooling power really well for larger values of the magnetic field ratio,  till the point where refrigeration is no longer possible.

\subsection{Experimentally realizable values of the cooling power}
To provide an estimate of the achievable cooling power, we assume that the cold reservoir is a normal metal, connected to the superconducting material through a thin insulating barrier, thus realizing a tunnel junction \cite{pekola_normal-metal-superconductor_2007}. The heat transfer to the magnetic field vortices occurs through quasiparticle tunneling through the junction. Assuming the area of contact to be $A$ and the width of the racetrack circulating fluxons to be $w$, the energy transported via the quasiparticles per unit time is \cite{manikandan2019superconducting, haack_efficient_2019} 
\begin{equation}
    P^\mathrm{qp}=\frac{2 A}{e^2 \mathcal R_\mathrm s}\int_0^{\infty}\mathrm dE\,E(f_\mathrm{FD}(T_\mathrm C)-f_\mathrm{FD}(T)).
\end{equation}
Here, $f_\mathrm{FD}$ refers to the Fermi-Dirac distribution and $\mathcal R_\mathrm s$ is the specific resistance of the junction. Given that fluxons are almost like small puddles of normal regions, we have approximated the junction transport to follow the normal-insulator-normal behavior. The above expression for small values of temperature difference $T-T_\mathrm C\ll (T+T_\mathrm C)/2$ evaluates to 
\begin{equation}
    P^\mathrm{qp}\approx\frac{\pi^2 k_\mathrm B^2 A}{3e^2 \mathcal R_\mathrm s}(T_\mathrm C-T)T_\mathrm C.
    \label{pqp}
\end{equation}
 Since the Fourier formula in Eq.~\eqref{Fourier} deals with unit volume heat transfer rate, comparing $P^\mathrm{qp}$ with $\dot{Q}\times w A$ leads to the   expression for $\kappa_{\mathrm{th}}$
\begin{equation}
    \kappa_{\mathrm{th}} = \frac{\pi^2 k_\mathrm B^2 T_\mathrm C}{3e^2 w \mathcal R_\mathrm s}.
    \label{kappa_expr}
\end{equation}
For numerical estimation, we look at the maximum cooling power predicted by expression \eqref{pcmax} 
\begin{equation}
    P_{\mathrm{cooling}}=P^{\max}_\mathrm C \times w A.
    \label{pcool}
\end{equation}
Now, assuming that the critical temperature of the superconductor is of the order of \unit{10}\kelvin, we assume $T_\mathrm H\sim T_\mathrm C=\unit1\kelvin$. Additionally, we assume $A=\unit1{\milli\meter\squared}$, $\mathcal R_\mathrm s=\unit2{\mega\ohm\usk\micro\meter\squared}$, $\phi_\mathrm C=0.8$, $\mu=0.5$, $l=1$, $\Omega =0.01$. Using Eqs.~\eqref{pcmax}, \eqref{kappa_expr} and \eqref{pcool} the maximum cooling power evaluates to be $P_{\mathrm{cooling}}\sim\unit{11}{\nano\watt}$ (or equivalently  $\unit{11}{\nano\watt\per\milli \meter\squared}$ of cooling power per unit area). 

\begin{figure}
    \centering
    \includegraphics[width=\linewidth]{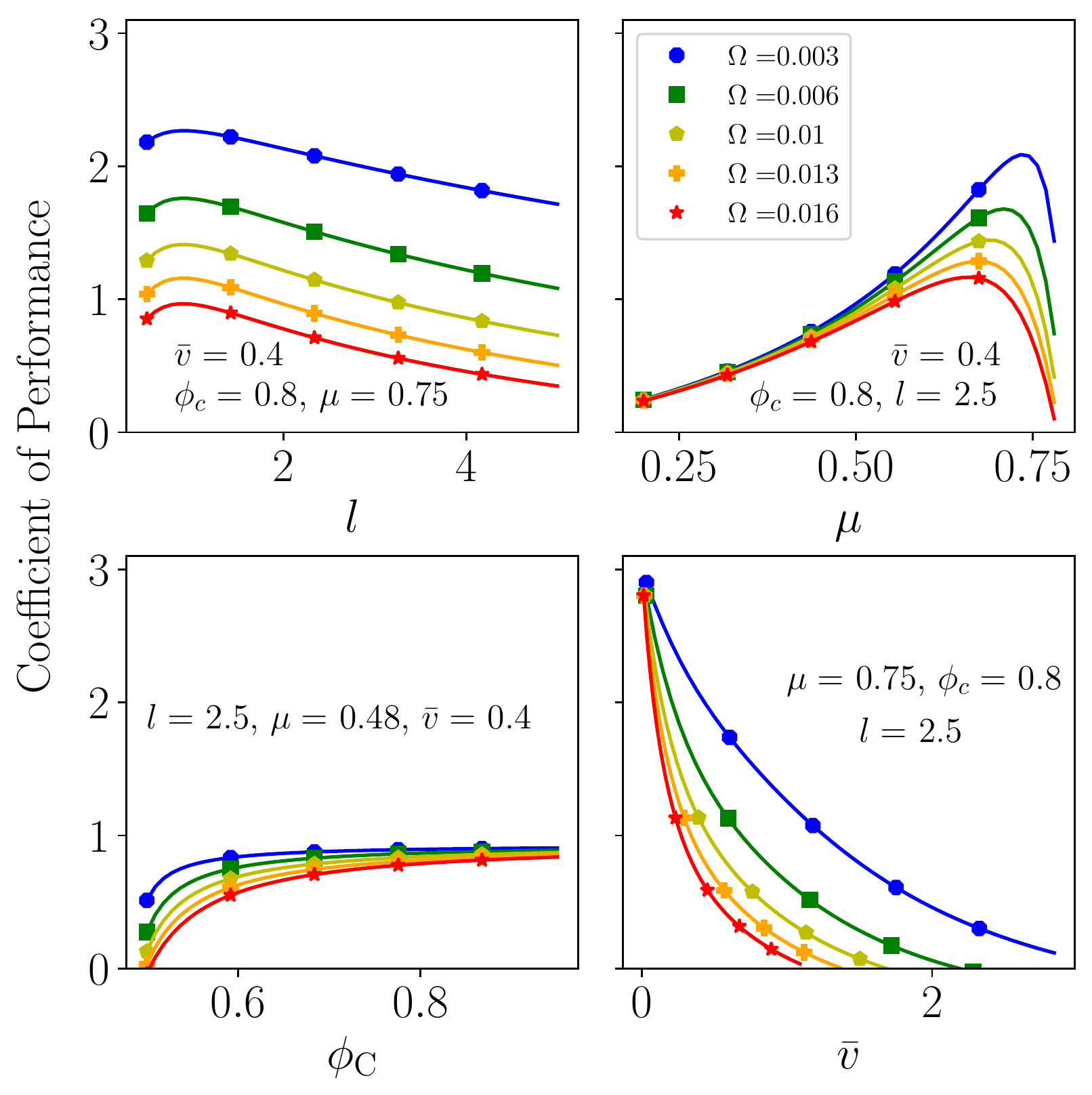}
    \caption{Plots of the coefficient of performance as a function of $l$ (top left), $\mu$ (top right), $\phi_\mathrm C$ (bottom left) and $\bar{v}$ (bottom right) for five different values of $\Omega$.  As one would anticipate, with decreasing dissipation (or $\Omega$), the coefficient of performance increases. From the top left plot, we see that the $COP$ first rises and then decreases as a function of length. We see a similar situation for the $COP$ as a function of $\mu$ in the top right plot. As expected with refrigeration, we see in  the bottom left plot  that with increasing temperature ratio, the $COP$ increases. It is worthwhile to mention that for the top two panels and the bottom right panel, the $COP_\mathrm {Carnot}$ is 4.}
    \label{cop_vs_params}
\end{figure}

\subsection{Coefficient of Performance }
With insights on the fluxon temperatures throughout the cycle and cooling power, we are now in  a position to characterize the refrigerator's performance.   To compute the coefficient of performance using the dimensionless parameters, we can write Eq.~\eqref{QH_dynamic} as 
\begin{equation}
\begin{split}
    &Q_{\mathrm H}=\gamma{ H}_{\mathrm R}T_{\mathrm H}^2\mu\left(l\bar{v}\Omega-(\tilde{\phi}_{\mathrm H}^2-\phi_{\mathrm L}^2)\right),\\=&\tfrac{Q_\mathrm C}{\mu}+\gamma{ H}_{\mathrm R}T_{\mathrm H}^2\bar{v}\Omega\left[l(\mu+1/\mu)+\tfrac{4}{3\mu}(1+\mu+\mu^2)\right].
\end{split}
    \label{Qh_simple}  
\end{equation}
The total work done $W=Q_\mathrm H-Q_\mathrm C$
\begin{equation}
    =\gamma{ H}_{\mathrm R}T_{\mathrm H}^2\left(l\bar{v}\Omega (1+\mu) -\mu(\tilde{\phi}_{\mathrm H}^2-\phi_{\mathrm L}^2)-(\tilde{\phi}_{\mathrm C}^2-\phi_{\mathrm R}^2)\right).
       \label{tot_w_diss}
\end{equation}
 Here, the first term corresponds to the work done by the battery to maintain the electrical current. The rest of the terms express the work done to move the fluxons along the magnetic field gradient. Interestingly, the temperatures satisfy
\begin{equation}
    \phi_{\mathrm L}^2+\tilde{\phi}_{\mathrm H}^2=\frac{1}{\mu^2}\left(\phi_{\mathrm R}^2+\tilde{\phi}_{\mathrm C}^2\right).
    \label{temp_identity}
\end{equation}

We can write
\begin{eqnarray}
    \frac{Q_\mathrm H}{Q_\mathrm C}=\frac{\mu\left( l\bar{v}\Omega-\left(\tilde{\phi}^2_\mathrm H-\phi^2_\mathrm L\right)\right)}{-l\bar{v}\Omega+\left(\tilde{\phi}^2_\mathrm C-\phi^2_\mathrm R\right)}.
    \label{qhqc_diss}
\end{eqnarray}
Using Eq.~\eqref{qhqc_diss} we can calculate the coefficient of performance as $COP=\tfrac{1}{Q_\mathrm H/Q_\mathrm C-1}$.

An important figure of merit for the refrigerator in view of achievable cooling is its coefficient of performance when the power output is maximum. For small fluxon speed, the coefficient of performance at maximum power (i.e.~$\bar{v}=\bar{v}_{lim,\mathrm R}/2$) turns out to be $COP(P_{\mathrm C}^{\text{max}})\simeq$
\begin{equation}
    \frac{\mu\left(l+\tfrac{2}{3}(1+\mu+\mu^2)\right)}{\left(l+\tfrac{2}{3}(1+\mu+\mu^2)\right)(3-\mu)-l(1-\mu^2)}.
    \label{copPcmax}
\end{equation}
Also,   for ~$\bar{v}\simeq\bar{v}_{\text{lim},\mathrm R}$, we  have  $P_\mathrm C=0$ and $Q_\mathrm C=0$. From Eq.~\eqref{Qh_simple} it is clear that, $Q_\mathrm H$ is not necessarily 0, leading to the fact that $COP(\bar{v}=\bar{v}_{\text{lim},\mathrm R})=0.$  Moreover, as $\bar{v}\to 0$,   $P_\mathrm C\to0$ and  $Q_\mathrm C$ is at a  maximum value shown in Eq.~\eqref{qcmax}. The heat transferred to the hot reservoir attains the minimum value $Q_{\mathrm H}^{\min}= \gamma H_\mathrm R T_{\mathrm H}^2 (\phi_{\mathrm  C}^2-\mu^2)/\mu. $ This leads to a maximum coefficient of performance $COP^{\max}=\tfrac{\mu}{1-\mu}.$ The highest value it can achieve is the Carnot coefficient of performance $\tfrac{\phi_\mathrm C}{1-\phi_\mathrm C}$ as $\mu\to\phi_\mathrm C$ and the cooling power goes to 0. 

We show the behavior of the coefficient of performance (calculated using Eq.~\eqref{qhqc_diss})  as a function of the dimensionless parameters $l$, $\mu$, $\phi_\mathrm C$, $\bar{v}$ and $\Omega$ in Fig.~\ref{cop_vs_params}. From the plots, it is clear that the dissipation, understandably,  causes a decrease in the coefficient of performance. For small lengths $l$, the fluxons do not get to interact with the cold reservoir for much time. Therefore, with increasing length, the heat withdrawn, and consequently, the coefficient of performance increases. However, as we keep increasing the length, $Q_\mathrm C$ reaches a limiting value but the work done by the current source keeps increasing. This decreases the coefficient of performance. This observation also explains the flattening of the curves with decreasing $\Omega$ since  the work needed to keep the current flowing also decreases. We see a similar behavior of the coefficient of performance as a function of the magnetic field ratio. As seen in  Fig.~\ref{cop_vs_mu0}, the $COP$ first rises as a function of $\mu$. However, as $\mu$ nears $\phi_\mathrm C$, the cooling power and therefore the $COP$ decreases. Next, as we increase $\phi_\mathrm C$, more heat can be exchanged with the cold reservoir, thus leading to an increase in the $COP$.   As $\phi_\mathrm C\to 1$, for small $\bar{v}$, the $COP$ tends to Eq.~\eqref{cop_sw}. Finally, it can be argued that with increasing speed,  the work done by the current source increases, and the heat withdrawn from the cold reservoir decreases, causing a significant decrease in the coefficient of performance.

\begin{figure}
    \centering
    \includegraphics[width=\linewidth]{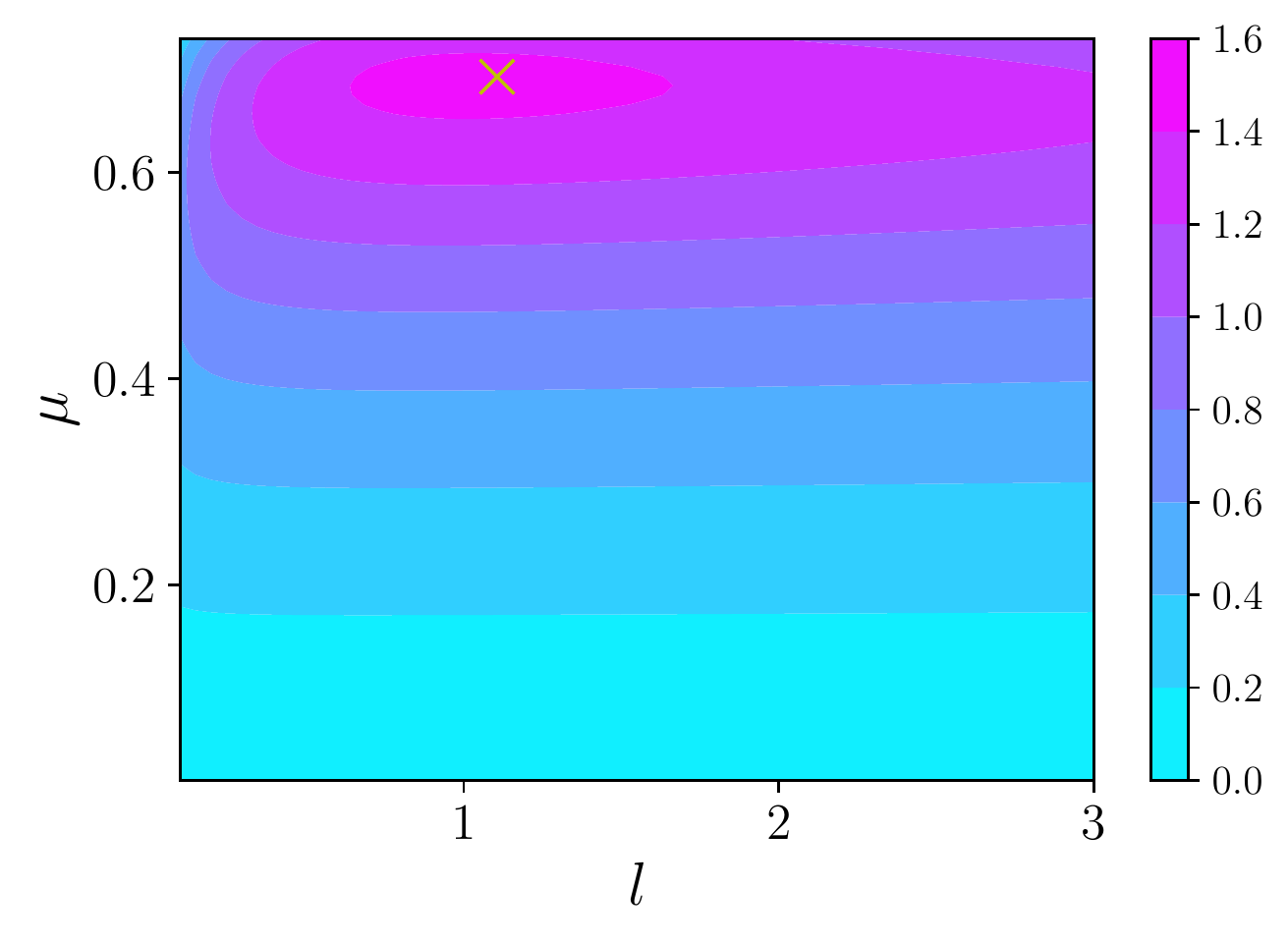}
    \caption{Plot of the coefficient of performance as a function of $\mu$ and $l$, for $\phi_\mathrm C=0.8$, $\bar{v}=0.5$, and $\Omega=0.01$. The yellow `$\times$' shows the maxima of the coefficient of performance calculated numerically. Due to dissipation, the maximum achievable $COP$ ($\sim 1.5$) is less than $COP_\mathrm{Carnot}$ ($=4$), while the cooling power at the maximum  $COP$ is nonzero ($\sim 0.06\kappa_\mathrm{th} T_{\mathrm H}/2 $).}
    \label{cop_vs_mu}
\end{figure}

These insights allude to the fact that for a given $\Omega$, $\phi_\mathrm C$ and $\bar{v}$, we can choose $\mu$ and $l$ that maximize the $COP$. Fig.~\ref{cop_vs_mu} shows the result of  numerical maximization of the $COP$---or minimization of Eq.~\eqref{qhqc_diss}---with respect to $\mu$ and $l$. Thus, for  fixed bath temperatures, current and superconducting material, this analysis can be used to  find an optimal value of the magnetic fields and the optimal geometry of the system.

\section{\label{discussions}Discussion}

Given that fluxons---which act like puddles of normal regions---predominantly carry the heat around the race-track cyclically, we considered a simple, Fourier law for heat transport across the interface for the steady state, where the heat current is proportional to the temperature difference across the interface, with the proportionality constant being the thermal conductivity of the interface. Such considerations, however, ignore different additional contributions to the heat transport possible in the system---originating from the energy level structure of the junction, or external voltage biases---and therefore have to be carefully accounted for to achieve the predicted cooling behavior in an experiment. For example, a small gap of the working superconductor can induce an above-the-gap quasi-particle transport from the hot to the cold reservoir across the sample, nullifying the cooling effect. Such quasi-particle transport across the working superconductor can be reduced by choosing either a superconductor of large-gap as the working superconductor, or by biasing the chemical potential of the reservoirs below the gap energy of the working superconductor. In this regime, only sub-gap transfer of charge is possible, via Andreev reflections~\cite{osti_4071988}, as the quasi-particle heat transport becomes hindered by the energy gap of the working superconductor.     Also, biasing the current source (which generates the circulation of fluxons) below the gap energy of the working superconductor further ensures that it will be a supercurrent (with no associated heat transfer) that generates the circulation, and not the above-the-gap quasi-particle current (which would add another heat contribution).  

Before we conclude, we wish to also comment on the role of phonons in the refrigeration process. The role of phonons is two-fold for the systems considered here. Presently, we have assumed that quasi-electrons and phonons decouple at low temperatures. Therefore one can ignore entropy exchanges between electrons and phonons in the transport of fluxons along the adiabatic arms of the cooling cycle. This results in simple linear cooling laws upon adiabatic magnetization, as we discussed in the present manuscript. On the other hand, when the interactions between quasi-electrons and phonons occur much faster than the adiabatic timescale, we note that the cooperative effect of phonons can actually help improve the cooling effect. This effect could be relevant for high$-T_{c}$, type-II superconductors, where one can operate the refrigerator cycles at higher temperatures. The second aspect is the effect of phonon-phonon interactions across a given interface. Such interactions results in a heat transport between phonon-phonon interfaces $i,j$ that is $\propto (T_{\text{ph}_{i}}^{4}-T_{\text{ph}_{j}}^{4})$ (known as the Kapitza coupling~\cite{pollack1969kapitza}). Although we have ignored the Kapitza coupling in our present work, assuming phonon mismatch across the relevant interfaces, we note that it could become relevant, for example when cooling down a substrate of phonons in stacked architectures.   

\section{\label{conclusions}Conclusions and outlook}
We presented a cyclic quantum refrigerator using magnetic field vortices in a type-II superconductor as the working substance. 
This design joins other mesoscopic quantum engines based on applied magnetic fields \cite{atalaya2012spintronics,sothmann2014quantum,sanchez2015chiral,manikandan2019superconducting,sanchez2015heat}.
The thermodynamic cycle consists of two adiabatic steps and two isothermal steps, where the superconductor is in contact with thermal reservoirs in the latter parts of the cycle. For both $s$-wave and $d$-wave models, we describe the thermodynamic quantities associated with the refrigerator and calculate the corresponding coefficient of performance. We also discuss the limiting cases where the thermodynamically permissible maximum (Carnot) coefficient of performance is achieved, while the cooling power generically drops to zero as the transport becomes reversible. We show that the $d$-wave case, under otherwise identical conditions, leads to a higher coefficient of performance. We also characterize the heat flow dynamics from the superconductors to the reservoirs and provide a transitory  description  of the temperature of the fluxons. As we incorporate dissipation into our narrative, we notice a decrease in the coefficient of performance, as expected

Our dynamical model dictates that  %as noted before, 
the velocity of the fluxons (and therefore the applied current) has to be small enough to allow the fluxons to exchange heat with the reservoir during the isothermal steps.  We also calculate the cooling power and show that it attains maxima for a certain value of the fluxon speed.   Additionally, we show that, depending on system parameters and applied fields, there is an optimal geometry of the system that allows the highest coefficient of performance.

 Our results point to several new future ventures one could undertake. Presently, we looked at the steady state dynamics of an ensemble of fluxons considering a mesoscopic sample. Their thermodynamics is well described by the fluxon gas limit as we discussed, and allows one to derive measurable figures of merit for the cyclic quantum refrigerator which can be tested in experiments. Beyond such explorations, it would also be exciting to explore the time-dependent, and transient behavior of the system considering a finite number of fluxons, as it might lead to new insights on cyclic refrigeration in the small quantum systems' regime of thermodynamics, where fluctuations---including quantum fluctuations---also become relevant~\cite{vinjanampathy2016quantum,kosloff2013quantum}; For example, making quantum mechanical observations of the dynamics of fluxons can incur a back-reaction on their dynamics, which could be harnessed to further improve the efficiencies of such microscopic devices~\cite{PhysRevResearch.4.033103,buffoni2019quantum,elouard2018efficient,elouard2017role}. We defer such analyses to future work.

 %Relevant \cite{https://doi.org/10.48550/arxiv.2112.03971}?  
\section{Acknowledgements}
We gratefully acknowledge 
  support from the U.S. Department of
Energy (DOE), Office of Science, Basic Energy Sciences
(BES), under Award No. DE-SC0017890. The work of SKM was supported by the Wallenberg Initiative on Networks and Quantum Information (WINQ). Nordita is partially supported by Nordforsk. We also thank Francesco Giazotto, Britton Plourde, Matthew LaHaye, Alok Singh, Bibek Bhandari and Sai Vijay Mocherla for helpful discussions.\\

\appendix
\section{Alternative  entropy gradient calculation for the dissipative case\label{appA}}
In this section, we provide an alternative dynamic strategy of including energy dissipation into our consideration. For the adiabatic processes stroke 1 and 2, heat dissipated in time $dt$ per unit volume at position $x$ is $dQ=n(x)\eta v^2 dt$. Now, consider an element of length $dx=vdt$. Change in entropy of the fluxons due to dissipation $TdS=dQ$. In time $dt$, fluxons at $x$ end up at $x+dx$. Therefore, if we express the unit volume entropy as a function of $x$ and $t$,
\begin{equation}
    S(x+dx,t+dt)=S(x,t)+\frac{n(x)\eta v}{T(x)}dx.
    \label{S_change}
\end{equation}
This leads to 
\begin{equation}
    \frac{\partial S}{\partial x}+v \frac{\partial S}{\partial t}=\frac{n(x)\eta v}{T(x)}.
    \label{S_eqn_gen}
\end{equation}
For the steady state this reduces to 
\begin{equation}
    \frac{\partial S}{\partial x}=\frac{n(x)\eta v}{ T(x)},
    \label{S_eqnA3}
\end{equation}
same as in \eqref{S_eqn}.

\bibliography{refs}
\end{document}